\documentclass{IEEEtran}

\usepackage{booktabs}
\usepackage{multirow} 
\usepackage{algorithmic}
\usepackage{graphicx}
\usepackage{multirow}
\usepackage{color,soul}
\usepackage{textcomp}
\usepackage{xcolor}
\usepackage{todonotes}
\usepackage{listings}
\usepackage{wrapfig}
\usepackage{tabularx}
\usepackage{url}
\usepackage{caption}
\usepackage{subcaption}
\usepackage{graphicx} 
\usepackage{tabularx}
\usepackage{beramono} 

\usepackage[framemethod=TikZ]{mdframed}
\definecolor{backgroundpage}{HTML}{F8F5F4}

\newenvironment{coloredframe}[2][]{
    \mdfsetup{
        skipabove=2pt, 
        hidealllines=true, leftline=true,      
        innerlinewidth=2pt, innerlinecolor=#2, 
        linewidth=0pt,
        backgroundcolor=#2!10
    }
    \begin{mdframed}}
    {\end{mdframed}}

\newcommand{\newtext}[1]{%
  \textcolor{black}{#1}
}

\newenvironment{keyfinding}[1][]{
    \mdfsetup{
        skipabove=5pt, 
        linecolor=blue,
        roundcorner=2pt,
        innerleftmargin=0.3cm,innerrightmargin=0.4cm, 
        linewidth=0.5pt,
        footnoteinside=false,backgroundcolor=blue!5}
    \begin{mdframed}}
    {\end{mdframed}}

\newcommand{\surveyquote}[2]{
    \begin{coloredframe}{darkgray}
    \vspace{5px}
    \small
    ``#2'' (#1).
    \vspace{5px}
    \end{coloredframe}
}

\definecolor{training}{HTML}{BD2222}
\definecolor{expert}{HTML}{A88332}
\definecolor{artifact}{HTML}{46A3A0}

\newenvironment{zeroindent}
  {\par\setlength{\parindent}{0pt}}
  {\par}

\newcommand{\pquote}[2]{\textcolor{#2}{\textbf{Participant #1:} }}
\newcommand{\gptquote}[1]{\textbf{ChatGPT:}}

\newcommand{\dialoguegpt}[2]{
    \begin{coloredframe}{#1}
    \vspace{5px}
    \small 
    \begin{zeroindent} #2 \end{zeroindent}
    \vspace{5px}
    \end{coloredframe}
}

\definecolor{p_pink}{HTML}{E0777C}
\definecolor{p_green}{HTML}{46A3A0}
\definecolor{p_blue}{HTML}{DBBA88}

\lstdefinestyle{chatPink}{
    basicstyle=\scriptsize\sffamily, 
    breaklines=true, 
    frame=leftline, 
    framerule=2pt, 
    rulecolor=\color{p_pink},
    rulesepcolor=\color{p_pink},
    showstringspaces=false,
    breakatwhitespace=true,
    escapeinside={(*@}{@*)},
    moredelim=[is][\color{p_pink}]{|}{|},
}
\lstdefinestyle{chatGreen}{
    basicstyle=\scriptsize\sffamily, 
    breaklines=true, 
    frame=leftline, 
    framerule=2pt, 
    rulecolor=\color{p_green},
    rulesepcolor=\color{p_green},
    showstringspaces=false,
    breakatwhitespace=true,
    escapeinside={(*@}{@*)},
    moredelim=[is][\color{p_green}]{|}{|},
}
\lstdefinestyle{chatBlue}{
    basicstyle=\scriptsize\sffamily, 
    breaklines=true, 
    frame=leftline, 
    framerule=2pt, 
    rulecolor=\color{p_blue},
    rulesepcolor=\color{p_blue},
    showstringspaces=false,
    breakatwhitespace=true,
    escapeinside={(*@}{@*)},
    moredelim=[is][\color{p_blue}]{|}{|},
}

\begin{document}


\title{Beyond Code Generation: An Observational Study of ChatGPT Usage in Software Engineering Practice}

\author{Ranim Khojah$^1$, Mazen Mohamad$^1$$^2$, Philipp Leitner$^1$, Francisco Gomes de Oliveira Neto$^1$\\
$^1$\textit{Chalmers University of Technology | University of Gothenburg} and $^2$\textit{RISE Research Institutes of Sweden}\\
Gothenburg, Sweden \\
khojah@chalmers.se, mazen.mohamad@ri.se, philipp.leitner@chalmers.se, francisco.gomes@cse.gu.se}





\maketitle

\begin{abstract}
Large Language Models (LLMs) are frequently discussed in academia and the general public as support tools for virtually any use case that relies on the production of text, including software engineering. Currently, there is much debate, but little empirical evidence, regarding the practical usefulness of LLM-based tools such as ChatGPT for engineers in industry. We conduct an observational study of 24 professional software engineers who have been using ChatGPT over a period of one week in their jobs, and qualitatively analyse their dialogues with the chatbot as well as their overall experience (as captured by an exit survey). We find that rather than expecting ChatGPT to generate ready-to-use software artifacts (e.g., code), practitioners more often use ChatGPT to receive guidance on how to solve their tasks or learn about a topic in more abstract terms. We also propose a theoretical framework for how the (i) purpose of the interaction, (ii) internal factors (e.g., the user's personality), and (iii) external factors (e.g., company policy) together shape the experience (in terms of perceived usefulness and trust). We envision that our framework can be used by future research to further the academic discussion on LLM usage by software engineering practitioners, and to serve as a reference point for the design of future empirical LLM research in this domain.
\end{abstract}


\section{Introduction}
With the advent of deep learning and Large Language Models (LLMs), artificial intelligence and machine learning have finally achieved widespread prominence, reaching far beyond academic or IT circles. Particularly, LLMs such as ChatGPT, Bard, or Co-Pilot are enthralling public perception with their power to generate human-competitive text. The potential for a tool that can generate correct text in an arbitrary format (including, for instance, programming language code) in virtually any context and based on very little input (the "prompt") seems almost unlimited.

However, this public prominence has also led to widespread criticism, in terms of both, ethical and practical concerns. Ethical concerns include aspects such as whether companies (e.g., OpenAI, Microsoft or Google) have the right to train a machine learning model based on data they have access to, but do not own. Practical concerns are more centered around the question to what extent an LLM, which fundamentally can only repeat and re-combine pre-existing text, can ever truly be helpful for creative tasks such as essay writing or coding --- particularly as current-generation LLMs are prone to "hallucinating" (generating plausible-sounding but inaccurate or downright non-sensical text) if their training data does not allow them to solve a specific task. 

Our goal in this paper is to shed light on the latter questions in the context of software engineering. We conduct an observational study of 24 professional software engineers working at 10 companies, who have used ChatGPT (GPT-3.5) for a period of one week in their daily tasks. We qualitatively analyse their chat protocols as well as their reflections after the study period, which they provide in form of an exit survey. Our ultimate study goal is to establish a framework for how software engineers use LLMs such as ChatGPT, and what factors influence their overall experience in terms of usefulness and trust in the generated advice or artifact. Particularly, we investigate the following research questions:\\

\noindent \emph{\textbf{RQ1}: For what kind of tasks do software engineers use an LLM-powered chatbot such as ChatGPT in their work?}

Our study \newtext{presents categories of ChatGPT usage, and} shows that software engineers used ChatGPT with three different high-level purposes in mind --- besides using it to directly manipulate artifacts (e.g., generate or fix code), they were also using it for learning as well as to get high-level guidance (which they then implemented themselves). Interestingly, we found that receiving guidance is the most common ChatGPT use case in our study. Most of the use cases that we observed were within the implementation phase of the software development lifecycle. However, our participants also used it for planning, design, and testing (and then often either as a tool for learning, or to brainstorm suggestions).\\

\noindent \emph{\textbf{RQ2}: What factors influence the personal experience of software engineering practitioners when using ChatGPT, particularly regarding perceived usefulness and trust?}

We found that a combination of internal and external factors influence practitioner's experience. In particular, we saw that the phrasing of prompts (particularly in terms of how much context is provided), and the participant's potential biases shaped their satisfaction with the generated results, and also how much they trusted the LLM. Additionally, some external factors, such as company policies and legal aspects, also influenced the experience of practitioners.

Approximately 75\% of participants found ChatGPT helpful to learn, and 50\% to reduce repetitive tasks. However, some participants reported a lack of trust in the results, and thoroughly double-checked any suggestions by ChatGPT. Particularly the lack of sources or references was seen as a challenge. Despite this, many practitioners still feel that ChatGPT was a useful tool that can make them more productive in their work.\\

We envision that our framework can be used by future research to further the academic discussion on LLM usage by practitioners, and to serve as a reference point for the design of future empirical LLM research in this domain. Therefore, we summarise the following contributions from our paper: 

\begin{itemize}
    \item We present a model that categorizes user purpose into three distinct types of interactions with ChatGPT.
    \item We introduce a theoretical framework with factors for assessing users' personal experiences when engaging with ChatGPT.
    \item We provide a re-analysis package encompassing quantitative data extracted from dialogues, including keywords used, dialogue classifications, as well as insights from exit surveys completed by practitioners, such as their levels of trust in ChatGPT.
    \item We outline a range of implications that summarize how ChatGPT has been utilized in software engineering tasks, shedding light on its practical applications and challenges in this context.
\end{itemize}
\section{Related work}
\label{sec:bg}

In existing work on bots in software engineering, the primary focus was on understanding the applications of bots \cite{santhanam2022bots} and setting expectations for the types of tasks and responsibilities that they should be able to do when assisting in development-related activities \cite{erlenhov2019current}. 
With the recent increased popularity of chatbots, the focus has shifted to studying LLMs, such as GPT-3.5 \cite{brown2020language} and Codex \cite{chen2021evaluating} (powering ChatGPT and Co-Pilot respectively). To the best of our knowledge, no taxonomy of different applications of such chatbots in software engineering has been proposed yet. However, possible abilities of ChatGPT have been suggested by Fraiwan and Khasawneh~\cite{fraiwan2023review}. LLM-powered chatbots have been studied in more specific contexts and software-related activities, for example code generation \cite{mastropaolo2023robustness, nguyen2022empirical, qian2023communicative}, requirements analysis \cite{ezzini2023ai}, test generation \cite{lemieux2023codamosa}, and others \cite{tufano2023automating, wood2018detecting, surameery2023use}.
In our work, we do not limit ourselves to a specific type of usage within software engineering, instead covering applications from all stages of the software development lifecycle.

An obvious question with LLM-based chatbots is to what extent they actually improve developer productivity. Productivity is notoriously hard to measure, as it depends on complex factors beyond the number of commits and pull requests. The SPACE framework \cite{forsgren2021space} presents different dimensions to consider when assessing a developer's productivity with respect to the developer's satisfaction, well-being, and communication within a team, in addition to the developer's activity, efficiency, and performance.
Meanwhile, Storey and Zagalsky~\cite{storey2016disrupting} provide a cognitive support framework to investigate the impact of different chatbots on the developer's productivity in terms of the chatbot's effectiveness and efficiency.
In our observational study, we use this framework to design our study material (i.e., the exit survey) and understand the impact of ChatGPT on our participants during their work. The advantage of Storey and Zagalsky's framework over SPACE is that it did not require us to measure productivity before and after the intervention.

Previous work has studied the helpfulness of chatbots in workflow-related tasks, such as work prioritization, scheduling, break reminders, and similar~\cite{kimani2019conversational}. Recent studies have been looking into the helpfulness of chatbots in terms of decision-making \cite{ahmad2023towards} and efficiency \cite{peng2023impact}, which can vary depending on the context. \newtext{ Waseem et al. }\cite{waseem2023using} \newtext{explored how ChatGPT can support undergraduate students in learning and improving software development skills.} In general, previous research has found that the experience of using a chatbot in software engineering can be effortful and frustrating \cite{weisz2022better}, \newtext{but also helpful in certain aspects like learning and simple software activities}\cite{waseem2023using}. However, there has been limited research on how chatbots can support engineers in core software engineering tasks, such as requirements engineering, testing, or coding.

Understanding the way software engineers interact with chatbots is important to have a better picture of what can impact productivity, effectiveness, and efficiency of chatbots. We are aware of one study that explored the types of interactions between developers and a code-generating chatbot (Co-Pilot) by Barke et al. \cite{barke2023grounded}. They introduce two interaction models of developers interacting with Co-Pilot, namely: acceleration mode, where developers use Co-Pilot to write things they already know faster, and exploration mode, where the developer explores solutions suggested by Co-Pilot.
We do not follow the same interaction models but rather explore our own, as our population is wider (not only developers working directly on code). Barke et al. also conducted their study primarily using students, whereas we exclusively rely on industrial practitioners using ChatGPT in their daily work.

\section{Methodology}
\label{sec:method}
\begin{figure*}[!ht]
    \centering
    \includegraphics[width=0.9\linewidth]{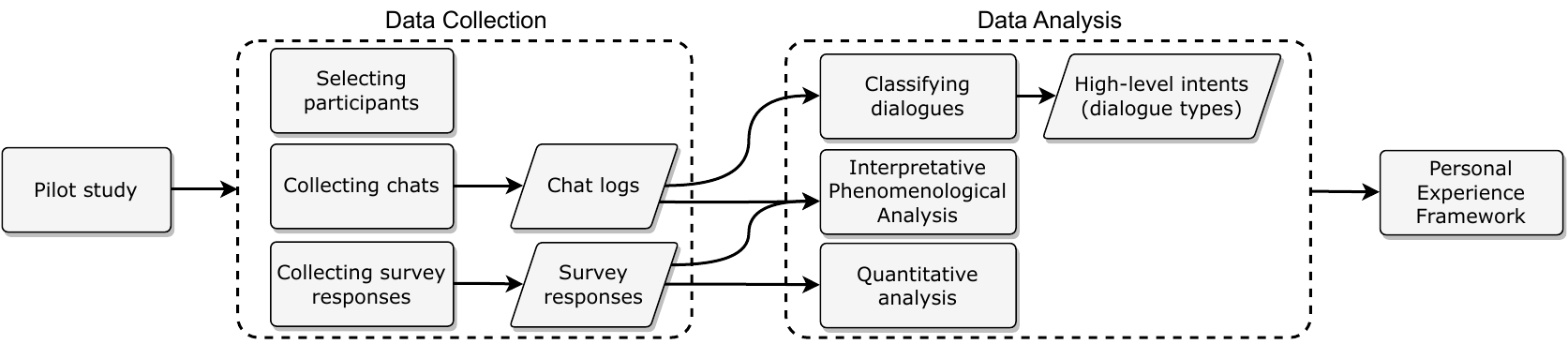}
    \caption{The main steps followed in our observational study.}
    \label{fig:method}
\end{figure*}

The goal of this study is to understand the personal experience of software engineers when interacting with ChatGPT.  It aims to identify the key factors that affect the interaction's usefulness and the level of trust users have in ChatGPT, consequently influencing the overall user experience. 
We perform an observational study as illustrated in Figure \ref{fig:method}. 
Initially, we conducted a pilot study to assess the feasibility of the research and refine the artifacts used with participants, such as training sessions, instructions about submitting their data, and an exit survey. Next, we detail those steps.

\subsection{Participants and data collection}

The data collection process consists of the \emph{(i)} selection of participants to join the study; and the \emph{(ii)} collection of their data. The participants interacted with ChatGPT (powered by GPT-3.5) over the course of 5 business days between March and July 2023 as part of their normal work. 
Then, \emph{(iii)} they sent us the logs of all chats and filled out an exit survey to capture their experience. Next, we detail how we selected participants and collected their data.

\textbf{Selecting participants: }
To initiate the study, we contacted 12 software organizations \newtext{of varying sizes in three counties in Europe,}inviting employees with diverse software engineering backgrounds. We conducted a brief training session introducing ChatGPT's architecture and demonstrating its capabilities and limitations in different domains (including software engineering) to ensure unbiased, domain-agnostic interactions.
During this session, attendees received a consent form to sign up for data collection, resulting in 42 registrations across 10 organizations \newtext{that expressed interest in participating in the study.}We follow our organisation's ethical guidelines ensuring participant anonymity, and allowing them to opt-out anytime during the study. Ultimately, \newtext{25 participants started the study out of which}24 participants completed the exit survey and shared their chat data. 
We present an overview of the participants in Table \ref{tab:participants-info} including their roles, work responsibilities, and domains. We classify the size of their organisation using the categories recommended by the European Commission \cite{euro-org-size}.

\begin{table*}[!ht]
    \centering
    \caption{Demographic information about the participants. We use IDs to refer to different participants and their corresponding organisations. The sizes used are Startup, Small and Medium enterprises (SME), and Large enterprises.}
    \label{tab:participants-info}
    \small
    \resizebox{2\columnwidth}{!}{\begin{tabular}{llllll}
    \toprule
    \textbf{ID} &\textbf{Role} & \textbf{Responsibilities} & \textbf{Org. ID} & \textbf{Org. Size} & \textbf{Domain}\\
    \midrule
    P1   & Software Tester  & UX inspections, usability and testing & A & SME  & Testing \\
    P2   & Test Engineer  & Test case execution & A &  SME  & Testing \\
    P3   & Test Engineer  & Test planning, design, and execution & A & SME  & Testing \\
    P4   &  Software Engineer  & Development and maintenance & B & SME  & E-learning \\
    P5   & Software Engineer  & Architecture and platform development &  B &  SME & E-learning\\
    P6   &  Full-stack Developer  & Development of web app's new features & B &  SME  & E-learning\\
    P7   & Product Manager  & Managing Frontend Digital Dep. & C &  Startup  & Medical \\
    P8   & Cloud Architect  & Architect applications and infrastructure &  C & Startup  & Medical \\
    P9   & Software Engineer  & Development and maintenance &   C & Startup &  Medical \\
    P10   & DevOps Engineer  & Manage and maintain the pipeline &  C & Startup  & Medical \\
    P11   & Software Developer  & Front-end development  & D &  Startup  & Gaming \\
    P12   & Game Developer  & Programming game logic &  E & Startup  & Gaming \\
    P13   & Software Developer  & Development, code review and testing & F &  Large  & E-commerce \\
    P14   & Group Manager  & Managing software development teams &  G &  Large  & Automotive \\
    P15   &  Software Engineer & Mobile app feature development & G &   Large  & Automotive \\
    P16   & Android Developer & Development and maintenance  & G &   Large  & Automotive \\
    P17   & System Leader  & System Design & G &   Large  & Automotive \\
    P18   & Sub-portfolio Manager  & Creation of roadmaps  & G &   Large  & Automotive \\
    P19   &  Product Manager &  Creation of roadmaps & G &   Large  & Automotive \\
    P20   & Android Developer &  Mobile app feature development  & G &   Large  & Automotive \\
    P21   &  Software Engineer & Development & H & Large  &   Consultancy \\
    P22   & Head of Operations & Define and establish processes  &  I &   SME & Consultancy \\
    P23   & Software Developer  &  Development and requirement analysis  & I &   SME  & Consultancy \\
    P24   &  Software Engineer &  Development and requirement analysis & J &   Large  & Automotive \\
    \bottomrule
\end{tabular}}
\end{table*}

\textbf{Chat Protocols: }
From 24 participants, we collected 130 chats in a web \newtext{archive}form \newtext{(.mhtml).} \newtext{The web archive format preserves the content of the web page (i.e., ChatGPT's web interface). This allowed us to obtain unalterable records of the chats and mitigate potential bias from the participants to remove parts of their chat logs.} For each participant, the chats are ordered by the date to better understand how the dialogues evolved throughout the study period.
We manually separate each chat into dialogues, where a dialogue is a series of consecutive prompts exchanged between the user and the chatbot on a \textit{similar topic or track}. This resulted in 208 dialogues that were then filtered to 180 dialogues related to software engineering. The excluded dialogues consisted of casual chit-chat or queries unrelated to software engineering.

\textbf{Exit Survey: }
Based on our research questions, we devised a survey to capture the overall experience and thoughts of the participants after completing the study. The survey covers five main elements: \emph{(i)} The overall experience; \emph{(ii)} specific scenarios where participants thought that their interaction went well (or not well); \emph{(iii)} feeling of trust in ChatGPT's answers; \emph{(iv)} perceived effectiveness and efficiency of ChatGPT; and \emph{(v)} learned lessons from their interactions and future usage of ChatGPT. The overall experience, the scenarios, and the takeaways were recorded in open-ended questions. Trust-related questions varied between Likert scale answers, multiple-choice, or open-ended questions. The questions about effectiveness focus on the feeling of achieving meaningful goals, whereas efficiency targets the feeling of ``doing things faster''. Those questions were guided by a cognitive support framework proposed to trigger reflections about the impact of bots in productivity \cite{storey2016disrupting}.

\subsection{Data analysis}

\textbf{Quantitative analysis:} We performed quantitative analyses on our data, consisting of descriptive statistics and visual analysis of the survey closed questions (e.g., diverging plots). We also aggregate the volume of dialogues based on the number of prompts per dialogue and how they evolve over time. This analysis aimed to identify inconsistencies in our data, such as identifying outliers (e.g., participants who interact substantially more often than others). In turn, we use qualitative analysis to investigate the content and nature of the dialogues. \newtext{Our analysis focuses on the flow of the dialogues and the overall user experience rather than the accuracy or the correctness of ChatGPT's responses.}

\textbf{Dialogue classification:} \newtext{The process of classifying the dialogues followed a protocol consisting of three main steps: 
The first step was done by three authors, and it involved (i) a pre-study that aimed to define the dialogue types that will be used for classification. The authors started with a literature review of existing dialogue types. Next, over three plenary meetings, the authors used the dialogue types from the literature to separately annotate a subset of our chat logs, compare the annotations, and discuss how the types can be adapted to software-related interactions. This process was repeated in every meeting until inconsistencies were resolved.}

\newtext{After the three meetings, the authors agreed on the dialogue types by Walton and Krabbe}\cite{walton2009burden}\newtext{, selected ``Deliberation'', ``Expert Consultation'', and "Didactic" dialogue types, and adapted their definitions to fit the nature of interaction with ChatGPT.}
One challenge was understanding the distinction between certain dialogue types, for example, delimiting between Expert consultation and Didactic when trying to characterize ``learning'' in a dialogue.

\begin{figure*}
    \centering
    \includegraphics[width=0.85\linewidth]{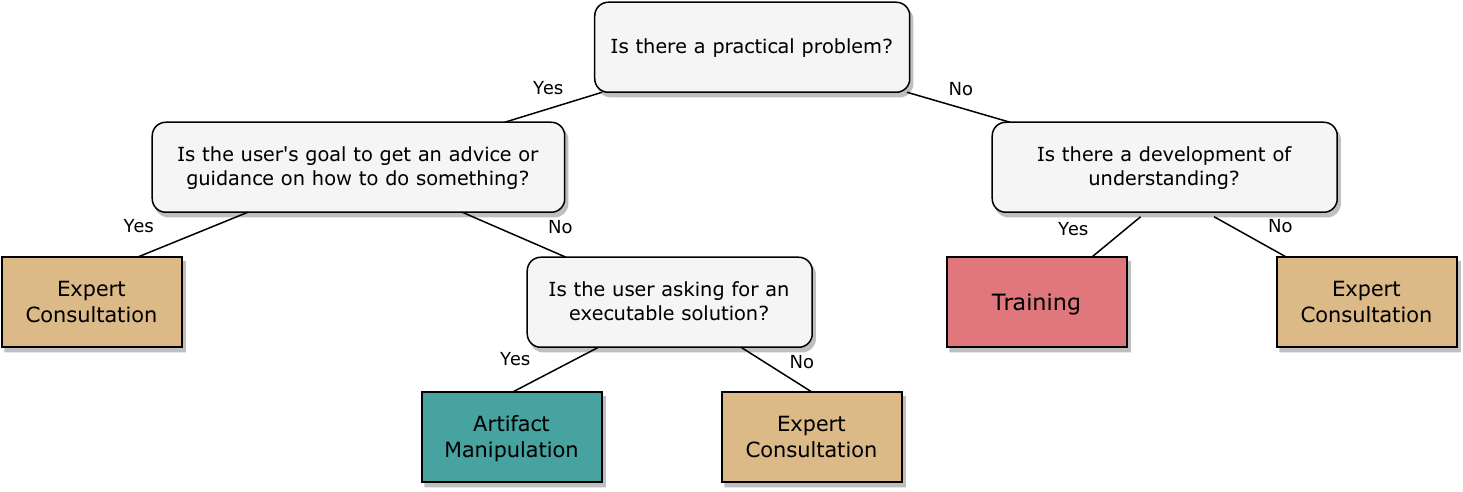}
    \caption{Decision tree to guide dialogue classification.\newtext{ The tree starts with determining if there is a practical problem. If yes, it checks if the user's goal is to be guided. If yes, it leads to Expert Consultation; if not, it checks if the user is looking for an executable solution leading to Artifact Manipulation or Expert Consultation. If there is no practical problem initially, it checks for a development of understanding in the dialogue leading either to Training or Expert Consultation.}}
    \label{fig:dialogue_tree}
\end{figure*}

\newtext{To address this challenge, we moved to the next step where the three authors (ii) compiled the outcome of the three meetings and annotation sessions and created a decision tree as a guide for classification (Figure }\ref{fig:dialogue_tree}\newtext{).}

\newtext{Using this schema, the first two authors had several sessions together and (iii)classified all of the dialogues at the same time. To better capture the semantics of the software engineering domain, we refer to dialogues in the ``Deliberation'' category as ``Artifact Manipulation'', and ``Didactic'' as ``Training''. We retained the term "Expert Consultation".}

\textbf{Interpretative phenomenological analysis:} 
Interpretative phenomenological analysis is a research method that examines personal lived experiences in-depth, emphasising understanding of the individual's own terms and seeking patterns across cases through interpretative processes~\cite{daSilva2015being, eatough2017interpretative}.
We performed the analysis of the chats and the survey in parallel in order to link the dialogues and their reported experience. The purpose of this analysis was to gain a better understanding of the overall experiences of the participants when interacting with ChatGPT. For each participant, we looked at the open-ended questions in the survey regarding \emph{(i)} the overall experience; \emph{(ii)} a scenario where ChatGPT was helpful; and \emph{(iii)} a scenario where ChatGPT was difficult to use. In parallel to that, we looked at the chats for each participant ordered from oldest to most recent in order to see how the dialogues evolved and to attempt to capture their thoughts and feelings during the interactions. For instance, we captured the frustration that was deducted from negative feedback to ChatGPT along with the description of the scenario in point \emph{(iii)} above. 
This allowed us to better understand the journey and mindset of different participants when interacting with ChatGPT.

\section{Findings}
\label{sec:results}
\begin{figure*}[h!]
    \centering
    \includegraphics[width=0.8\linewidth]{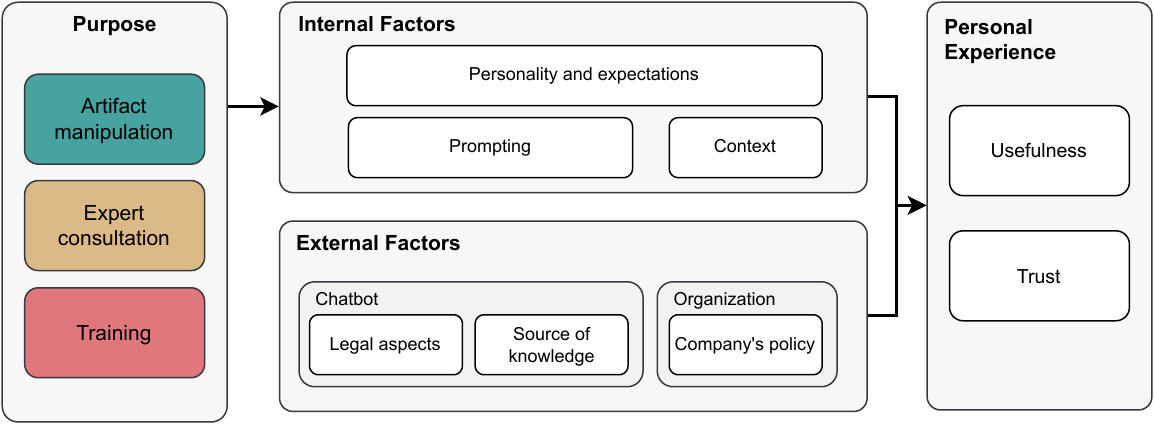}
    \caption{A theoretical framework of the factors that influence the personal experience of interactions with ChatGPT in industrial software engineering.}
    \label{fig:framework}

\end{figure*}

In this section, we introduce a theoretical framework, shown in Figure~\ref{fig:framework}, to assess the personal experience of an interaction with ChatGPT. This framework is based on a qualitative analysis of the dialogues that our participants had with ChatGPT, as well as on the analysis of the exit surveys submitted by our participants. A \textit{dialogue} (or interaction), in this context, is meant to include an initial prompt (and answer) as well as potential follow-ups triggered by the user.

At the core of the framework is the user's high-level \textbf{purpose} with the interaction. Based on our dialogue classification scheme (Figure~\ref{fig:dialogue_tree}), we identify three distinct dialogue types (artifact manipulation, expert consultation, and training), which fundamentally impact a user's experience. Additionally, we find three (user) \textbf{internal factors}, which include how the prompt is phrased and the contextual information provided in the prompt, as well as the user's personality and expectations. 
Orthogonally, we identify three \textbf{external factors} (i.e., factors \textit{not} directly related to the user or purpose), which include legal aspects, such as OpenAI's data policy on how the prompts are stored and used, ChatGPT's source of knowledge (the current model that powers ChatGPT has limited knowledge until 2021), and the user's company's policies with regard to the usage of ChatGPT or generative AI in general. Purpose, internal, and external factors together steer the \textbf{personal experience} of the user, which we break down into two concrete elements: the perceived usefulness of the interaction, and how much the user trusts the chatbot. Next, we detail the four main elements in our framework.

\subsection{Purpose}
\label{sec:personas}

Our study participants primarily use ChatGPT in one of two fundamental ways: (i) for artifact manipulation, i.e., in a goal-oriented manner with the expectation that ChatGPT will produce or modify a concrete solution or artifact, such as program code; or (ii) for information seeking. We further distinguish two subtypes of information seeking, namely expert consultation, where the user seeks problem-specific guidance, but not necessarily a tangible solution, or for training, where the user's goal is predominantly to \textit{learn} rather than to solve a specific problem. Most of our participants showed varied purposes when chatting with ChatGPT, but many had a predominant usage pattern, allowing us to identify keyword patterns and user reactions to ChatGPT's responses.
Therefore, those different purposes directly influence many of the users' internal factors, such as how the prompt is phrased, as well as the perceived usefulness of the interaction and trust that users have in ChatGPT.

\begin{table}[!ht]
\centering
\caption{Descriptive statistics of the purpose of ChatGPT interactions, such as the mean length of dialogues and its standard deviation (SD). The column ``Participants'' shows how many participants had \textit{at least one} interaction with this high-level purpose during the observation period.}
\label{tab:dialogtypes}
\begin{tabularx}{\linewidth}{Xrrcr}

\toprule
\textbf{Purpose} & \textbf{Interactions} & \textbf{Participants} & \textbf{Avg. length} & \textbf{SD} \\
\midrule
Artifact Manipulation & 57 (31.7\%) & 17 (70.8\%) & 3.12 & 3.86 \\
Expert Consultation & 112 (62.2\%) & 22 (91.6\%) & 2.47 & 2.60  \\
Training & 11 (6.1\%) & 5 (20\%)  & 4.00 & 1.84 \\
\bottomrule
\end{tabularx}
\end{table}

Table~\ref{tab:dialogtypes} lists how frequent the three high-level purposes were among our study population. The column "Participants" lists how many participants used ChatGPT with this purpose at least once. The table also reports on average length of interactions in number of prompts, and the standard deviation of this length. Expert consultations are the most common use of ChatGPT (62\% of dialogues and used at least once by all but two of our participants), followed by artifact manipulation. The training purpose was less common. This is unsurprising given our study's focus on workplace usage, where many sought assistance with specific work tasks rather than engaging in broader learning. However, it's worth noting that 20\% of participants (5 out of 24) engaged in task-independent learning. Additionally, we observe that training-focused dialogues tend to be longer than the ones directed at a specific work task. Finally, the significant variation in dialogue length arises from how each participant interacts with ChatGPT. For example, some participants consistently ask follow-up questions, while others abruptly end their conversations. Training dialogues exhibit less variation, indicating a more consistent discourse in these interactions.

\begin{table*}[!ht]
\centering
\caption{Number of dialogues per purpose and stage in the software development lifecycle}
\begin{tabularx}{0.75\linewidth}{llllll}
\toprule
\textbf{Purpose} & \textbf{Unspecific} & \textbf{Planning and Analysis} & \textbf{Design} & \textbf{Implementation} & \textbf{Testing} \\
\midrule
Artifact Manipulation & 12 & 6 &  1 & 34  & 4 \\
Expert Consultation  & 21 & 5 &  5 &  69 & 12 \\
Training  & 2 &  3  & 2  & 1  & 3 \\
\hline
\textbf{Total} & 35 (19.4\%) & 14 (7.8\%) & 8 (4.4\%) & 104 (57.8\%) & 19 (10.6\%) \\
\bottomrule
\end{tabularx}
\label{tab:sdlc-count}
\end{table*}

Table~\ref{tab:sdlc-count} displays when in the software development lifecycle ChatGPT was used by our participants. Unsurprisingly, most interactions (104, or about 58\%) were during implementation. However, we have seen usage of the tool in all phases of the lifecycle. Additionally, approximately 20\% of dialogues were not clearly associated to any specific phase in the software development lifecycle (e.g., side tasks, see Section~\ref{sec:manipulation} below).

\begin{keyfinding}
\textbf{Finding:} Each participant showed varied intents throughout their week using ChatGPT. Most of our participants are looking for more conceptual guidance rather than concrete solutions (e.g., code or test cases) from ChatGPT. 
Most interactions with ChatGPT are in the implementation phase, however, our participants have also used the tool in all other phases of the software development lifecycle.
\end{keyfinding}

We now discuss the three high-level purposes in more detail. Figure~\ref{fig:purpose} schematically depicts the subtypes we identified for each high-level purpose.

\subsubsection{Artifact Manipulation}
\label{sec:manipulation}
Artifact manipulation dialogues are characterized by the use of imperative forms to ask ChatGPT to perform an action or a task e.g., generate, refactor, fix, etc.
Artifact manipulation represents about one third of all interactions in the study period.

Artifact manipulation interactions are usually short --- users seem to either receive the results they are expecting quickly or give up. However, the dialogue was longer in a few cases where users were persistently trying (and failing) to get ChatGPT to provide a source for their answer. If users provide feedback to ChatGPT during an artifact manipulation interaction, it is usually negative, indicating problems with the generated solution. Concrete tasks are usually, but not exclusively, implementation-related, arguably because generating concrete code is a fairly obvious use case for generative AI (see also Table \ref{tab:sdlc-count}), as well as due to the high ratio of software developers among our study population (see also Table~\ref{tab:participants-info}). 

\dialoguegpt{artifact}{

\pquote{21}{artifact} "Rewrite this in functional Java 8: [code]?"

}

\begin{figure*}[!ht]
    \vspace{5px}
    \centering
    \includegraphics[width=\linewidth]{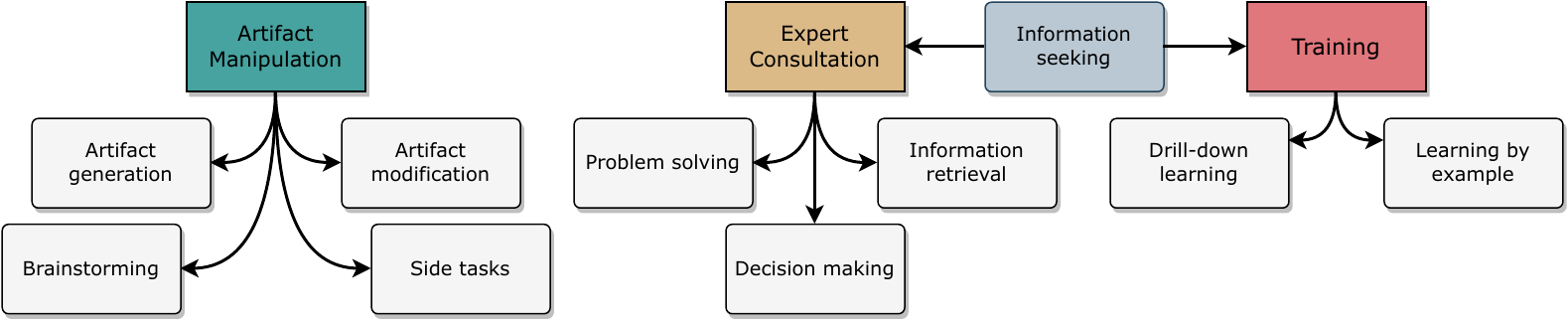}
    \caption{Taxonomy of purposes for the usage of ChatGPT in software engineering.}
    \label{fig:purpose}
\end{figure*}

We have identified four types of artifact manipulation: (i) generating artifacts, (ii) modifying artifacts, (iii) brainstorming, and (iv) handling side tasks (see also Figure~\ref{fig:purpose}). While all four involve creating software artifacts at different stages of software development, they vary in the degree of creativity and originality expected in ChatGPT's responses.

Participants aiming for \textbf{artifact generation} used ChatGPT to \textit{create} various types of artifacts such as code, software architecture, test cases, and even development processes, from scratch. Typically, participants provide descriptions or constraints as to what should be created. 
In contrast, participants wishing for \textbf{artifact modification} often provide an existing artifact to be improved, refactored, or even fixed. For instance, we saw that bug-fixing requests to ChatGPT are usually accompanied by an error message, whereas refactoring requests come with a requirement (e.g., call certain libraries). Below, the dialogues from Participant 7 (generation) and Participant 6 (modification) contrast both types of purpose.

\dialoguegpt{artifact}{

\pquote{7}{artifact} "Create a full agile process for a 2-member development team (a senior backend developer and a junior full stack developer). What meetings are needed every week, 2 weeks and month?"

\textbf{ChatGPT:} "Here is an Agile process that could be followed ..."
}

\dialoguegpt{artifact}{
\pquote{6}{artifact} "The following code snippet is giving me an error [code] [error] Can you fix it"

\textbf{ChatGPT:} "Yes, the error indicates that [...] To fix the error ... [code]"

\pquote{6}{artifact} "The [function name] function that you called does not exist"

\textbf{ChatGPT:} "I apologize for the confusion. You are correct ... [corrected code]"
}

Note that, while artifact generation requires specifications of what to generate, artifact modification requires providing the actual artifact to modify (in addition to some requirements of what to change). This can be problematic in a work context, as sharing production code with third parties may be against company policy. 

On the other hand, we have observed that artifact manipulation requests do not depend so much on providing extensive "additional context" (e.g., other parts of the system, dependencies, etc.), which is often required to generate correct, working code artifacts from scratch. This indicates that there may be room for more integrated generative AI tools, such as Co-Pilot, to support artifact generation. 
    
\textbf{Brainstorming} is important in software engineering, as it yields original ideas used for requirements elicitation, architectural solutions, or software innovation. Notably, the goal of brainstorming in this sense is to generate multiple alternative working artifacts, not just ideas or a single ideal solution. Some of our study participants (e.g., Participant 18) used ChatGPT as a brainstorming engine in this sense, for example, to formulate concrete user stories in the planning and analysis stages of their projects.  Finally, participants also utilized ChatGPT to streamline potentially time-consuming or labor-intensive \textbf{side tasks} (e.g., Participant 1). These tasks encompass activities like file organization, refining documentation wording, and more.

\dialoguegpt{artifact}{
\pquote{18}{artifact} "Can you brainstorm user stories for a [type of user] wanting to connect two phones to a [software system]"

\textbf{ChatGPT:} Certainly, here are a few user stories ...
}

\dialoguegpt{artifact}{

\pquote{1}{artifact} "Could you make 20 codes containing 4 numbers and 2 letters?"

\textbf{ChatGPT:} "Sure, here are 20 codes consisting of 4 numbers and 2 letters: ..."

}  

The use cases above illustrate the nuances in the participant's experience when using ChatGPT to manipulate artifacts, indicating that LLM usage is significantly broader than pure code generation even within the artifact generation purpose. We have seen examples ranging from fixing existing code to inspiration-focused dialogues intended to originate ideas. The responses to our exit survey also show the impact of those varying expectations on the participants' perceived usefulness of ChatGPT.

\begin{keyfinding}
\textbf{Findings:} Participants employed ChatGPT for artifact manipulation, ranging from practical solutions rooted in existing artifacts to idea generation through inspiration-focused dialogues.
\end{keyfinding}

\subsubsection{Expert Consultation}
\label{sec:expert}
Seeking expert consultation is, by far, the most common purpose of interactions in our study (62\%). Our participants often ask for resources to assist them in their work tasks, such as instructions, advice, or detailed information. Unlike the artifact manipulation purpose, the goal of these interactions is \textit{not} to obtain a concrete solution, but rather a nudge in the right direction. In that sense, our participants often utilize ChatGPT as a "virtual colleague" they can turn to for high-level advice or as a more expedient alternative to searching the Internet.
\vspace{5px}

\dialoguegpt{expert}{
\pquote{12}{expert} "What is a good data structure to use for [use case]?"
}

Expert consultation dialogues are often short. Our participants prefer ``how to'' questions for this purpose, hence indicating the desire for \textit{guidance towards} a solution rather than a ready-to-use artifact. This also aligns with existing work~\cite{melo2020exploring}, which also observed that developers are interested in being guided in their tasks rather than having a chatbot that can execute tasks on their behalf. 
\vspace{5px}

\dialoguegpt{expert}{
\pquote{21}{expert} How do I read the following [violation]?
}

The prompts in our study revealed an underlying concern to understand the steps required to reach a solution. This is an important distinction to the training purpose, where follow-up prompts reveal the participant's desire for learning. We observed that expert consultation interactions have three main purposes: (i) solve problems, (ii) retrieve information, and (iii) make decisions (see also Figure~\ref{fig:purpose}).

Practitioners aiming to \textbf{solve problems} utilized ChatGPT to get instructions on implementing solutions, rather than expecting a concrete solution artifact (e.g., Participant 6). Alternatively, participants use ChatGPT to \textbf{retrieve information} usually found in discussion forums, or long articles. Participants found querying ChatGPT to be a more time-efficient alternative to reading these primary sources, especially when seeking commands for tools such as git or clarifying the syntax of programming APIs. Some participants (e.g., Participant 14) also used ChatGPT to retrieve basic facts or definitions of software engineering concepts, techniques, and technologies.

\vspace{5px}

\dialoguegpt{expert}{
 \pquote{6}{expert} "How can I perform an action in a web application in only one tab [while] multiple tabs [are] open"
 
 \textbf{ChatGPT:} "Here's how you can perform an action in a specific tab: [steps to follow]"
}

\dialoguegpt{expert}{
    \pquote{14}{expert} "what is kanban?"
    
    \textbf{ChatGPT:} "Kanban is a visual project management tool and methodology that ..."
}

Many practitioners also sought ChatGPT's recommendations for \textbf{decision-making}. In these interactions, ChatGPT would either provide a single recommendation upon request or offer multiple suggestions. While some preferred multiple recommendations, many participants would subsequently ask ChatGPT to select the most suitable recommendation for their context.
\vspace{5px}

\dialoguegpt{expert}{
    
    \pquote{3}{expert} "Give me example of a good test approach for API testing"
    
    \textbf{ChatGPT:} "A good test approach for API testing typically involves a combination of manual and automated testing [...] This can be done using tools such [...]"
    
    \pquote{3}{expert} "What programming languages are most suitable for those test tools that are listed?"

}

These use cases are interesting, in that ChatGPT mostly provides a, potentially faster, alternative to reading or searching the Internet. That is, ChatGPT does not allow our participants to solve a problem that they otherwise would not have been able to solve, but potentially provides a more productive solution. This is similar to how existing research has shown bots in software engineering to be used~\cite{erlenhov2020empirical}.

\begin{keyfinding}
\textbf{Findings:} The most common use case for ChatGPT is getting expert guidance on a specific topic. In this sense the tool can serve as a, potentially more productive, alternative to asking a colleague or searching the Internet.
\end{keyfinding}

\subsubsection{Training}
\label{sec:training}
Lastly, the \textbf{training} purpose differs from expert consultation as participants are not seeking a direct solution to a practical issue. Instead, they aim to acquire broader theoretical or practical knowledge related to a work task. Although these interactions are relatively infrequent (approximately 6\% of all interactions), they typically result in longer conversations featuring multiple follow-up queries to clarify previous answers. Participants who seek to learn often confirm their comprehension of ChatGPT's responses, request further explanations or examples, or pose additional questions to expand the scope of their learning.

\dialoguegpt{training}{
\pquote{1}{training} "describe API as simple as you can"
    
\textbf{ChatGPT:} "API stands for Application Programming Interface. It is a way ..."
    
\pquote{1}{training} "And how do I test API?"
    
\textbf{ChatGPT:} "There are many tools available, such as Postman, SoapUI, or Insomnia"
    
\pquote{1}{training} "Postman or SoapUI?"
    
\textbf{ChatGPT:} "Here is an overview ..."
    
\pquote{1}{training} "how to write tests in postman?"
}

Dialogues showing a training intent often show a progression from broader initial questions to increasingly detailed ones. This reflects how participants' understanding of the subject matter grows through the chatbot's responses. Dialogues with a training purpose are diverse; we found them in dialogues that belong to all different stages in the software development life-cycle, with fewer implementation-related questions than for other purposes (Table~\ref{tab:sdlc-count}).

We observe that participants employ two distinct learning strategies when using ChatGPT (Figure~\ref{fig:purpose}): (i) drill-down learning and (ii) learning by example. In \textbf{drill-down learning}, participants typically initiated the conversation with a general prompt or a request for a relatively abstract piece of information or definition. Subsequently, they would seek additional details or explanations related to the information provided in the response.

\dialoguegpt{training}{
    \pquote{9}{training} "How to automate testing?"
    
    \textbf{ChatGPT:} "Automating testing can help improving [...]. Here are some general steps you can follow 1. [...]. 2. Select a testing framework [...]"
    
    \pquote{9}{training} "What are testing frameworks?"
    
    \textbf{ChatGPT:} "Testing frameworks are software tools or libraries that [...]. Some popular testing frameworks include JUnit, Selenium ..."
    
    \pquote{9}{training} "How can I configure JUnit?"
    
    \textbf{ChatGPT:} "Configuring JUnit involves ..."
}

On the other hand, participants adopting the \textbf{learning by example} approach generalize concepts by examining concrete cases that illustrate those concepts. Unlike the previous method, which began with abstract concepts and moved towards specific solutions, the example-based approach allows participants to explore a wide range of related concepts rather than delving deeply into definitions. The following example demonstrates how Participant 18 initially inquired about a specific artifact but ended up showing curiosity about the components that make up the desired artifact.
\vspace{5px}

\dialoguegpt{training}{
\pquote{18}{training}"What belongs to a software roadmap?"

\textbf{ChatGPT:} "Typically, a software roadmap includes the following elements: 1. [...] 2. Vision [...] 4. Release Schedule [...] 7. Metrics: Key Performance Indicators (KPIs) that will be used to measure the success of the software project. 8. [...]"

\pquote{18}{training} "can you give me some good example of metrics to measure the success of the software project?"

\textbf{ChatGPT:} "There are many metrics that depend on the goal of the software project [...]. Here are some metrics that are commonly used [...]."

\pquote{18}{training} "Can you give some examples of goals for a software roadmap?"

\textbf{ChatGPT:} "Some examples of goals for a software roadmap might include: ..."

\pquote{18}{training} "Can you give some examples of vision for a software roadmap?"

\textbf{ChatGPT:} "Some examples of vision for a software roadmap might include: [...] The vision should help guide the development of key features and priorities."

\pquote{18}{training} "Can you give examples of key features?"

\textbf{ChatGPT:} "Key features ..."

\pquote{18}{training} "Can you give some examples of release schedule?"

\textbf{ChatGPT:} "Some examples of release schedule for a software roadmap might include: ..."
}

\vspace{5px}
One key challenge in analysing the participants' dialogues was to distinguish between the intent of training and expert consultation. However, participants who engaged in training interactions reported a more consistent perception of ``learning something useful'' from ChatGPT compared to those using it as an expert consultant in the exit survey.

\begin{keyfinding}
\textbf{Findings:} Dialogues with a training purpose are infrequent, but often longer than other interactions. Participants use ChatGPT predominantly to learn in phases other than the implementation phase.
\end{keyfinding}

\subsection{Internal Factors}
\label{sec:internal}

Internal factors are aspects of the chatbot user, such as their personality or how they prompt ChatGPT. These factors, along with the purpose, directly influence the perceived quality of the outcome of the interaction.

\subsubsection{Prompts}
\label{sec:prompt}

Participants engage with ChatGPT through prompts, and the way they structure these prompts affects ChatGPT's responses. We analyzed prompt structure in terms of linguistic structure (e.g., grammar and morphology) and prompt content, such as including snippets of code to be fixed. Although the majority of users constructed grammatically correct and coherent sentences, some simply combined keywords without using conjunctions or verb inflections when interacting with ChatGPT, resembling a search engine query (e.g., Participant 11).
In their exit survey, some participants (e.g., Participant 2) even drew comparisons between their interactions with ChatGPT and using Google search.

\dialoguegpt{expert}{
    \pquote{11}{expert} "align text left blender python"
    
    \textbf{ChatGPT:} "To align text left in Blender using Python, you can use ..."
}

\surveyquote{Participant 2}{I found that the answers were easier to follow than the ones I got from Google.}

We found no disparity in perceived usefulness between participants who employed keywords and those who formulated correct sentences, even though the latter approach yielded more accurate answers from ChatGPT. Instead, we observed that the presence or absence of project-specific information, referred to as contextual information or simply \textbf{context}, had a more significant influence on the participants' overall experiences. 

This contextual information includes project-related details such as product requirements, code snippets, domain-specific constraints, and process structure, shared by users during their interactions. It should be noted that some participants had privacy concerns (or, relatedly, concerns about whether sharing information with ChatGPT would be allowed according to their company's policy) and hence opted not to provide sufficient purpose when asking domain-specific or technical questions.

In expert consultation, context is less commonly provided (and often not as necessary, as users are not expecting a directly usable artifact). If context is provided in these interactions, it is often in dialogues that aim to explain a piece of code or an error message (where the relevant artifact is provided), or decision making. Participants sometimes specify a role or persona for ChatGPT. We assert that such prompt structure (``As a X, do Y'') is employed by some participants because it aligns with the common phrasing of ChatGPT examples found on the Internet. We illustrate both types of context in the dialogues below, where Participant 7 provides project-specific conditions, and a role, that ChatGPT should consider in its answer.

\dialoguegpt{expert}{
\pquote{7}{expert} "what is the best SaaS solution for the [global area] from a logistics perspective?"
}

\dialoguegpt{expert}{
    \pquote{7}{expert} "As a software engineer, to what extent would Smoke testing help expose vulnerabilities?"
}

Our data indicates that the purpose behind a dialogue strongly influences the amount and type of context participants provide. For example, in training interactions, participants naturally provide less context as these dialogues with ChatGPT are exploratory in nature. When asking ChatGPT to generate artifacts, participants rarely included sufficient context in their prompts (in the form of requirements or applicable constraints on the solution). In contrast, modifying an artifact implies providing it and often results in improvements in perceived usefulness.
Additionally, we noticed that participants often interrupted their dialogues shortly after  telling ChatGPT that the answer did not align with the provided contextual information (e.g., code with compilation errors), indicating that these users have given up on the interaction (or ChatGPT in general).

\begin{keyfinding}
    \textbf{Findings:} Many participants struggle with providing sufficient context, either because they do not know what context to provide or because providing it may violate company policy. These users sometimes end up frustrated at these perceived failings of ChatGPT, terminating interactions mid-way.
\end{keyfinding}

\subsubsection{Personality and expectations}
\label{sec:expectation}

Other important factors influencing the participants' personal experience are their opinion of AI in software engineering, how much they know about generative AI, and their tolerance to inaccuracies in responses. Some participants are optimistic and aware of ChatGPT's limitations, hence expecting occasional errors and inaccuracies. Those participants focus more on ChatGPT's strengths and adapt their usage accordingly. In their exit survey, participants who use the tool with a training intent express positive sentiments more often (e.g., Participant 22).

\surveyquote{Participant 22}{I am still discovering its powerful ability to generate information. The speed in which it does that and the human-like attentiveness are astonishing.}

When participants without prior experience with ChatGPT approached the tool, they usually initially tested the waters by exploring questions unrelated to their work, or via questions that they already know the answer to. Oftentimes, these users expect complete and fault-free responses, and end up judging ChatGPT's answers as unsatisfactory (e.g., Participant 16). We also observed that many participants asked for sources, which ChatGPT cannot accurately provide by construction\footnote{\url{https://www.microsoft.com/en-us/microsoft-365-life-hacks/writing/using-chatgpt-for-source-citation}}, leading to \textit{scepticism} regarding ChatGPT's answers. Sceptical participants also often avoid adding necessary context so as to not share potentially sensitive information.

\surveyquote{Participant 16}{Anyone who lacks knowledge of the core basics and principles of the subject topic should be very discouraged from using [ChatGPT].}

\begin{keyfinding}
\textbf{Findings:} Sceptical participants are often confirmed in their believe that ChatGPT is not useful for their work (e.g., because of inaccurate responses to test queries), and stop using it quickly. Participants that use ChatGPT for training are often more happy with the results they receive.
\end{keyfinding}

\subsection{External Factors}
\label{sec:external}
We identified three external factors influencing our participants' experiences. Firstly, some participants raised legal and ethical concerns, particularly regarding the data used to train the AI and the storage and utilization of prompts, which sometimes contain sensitive information. 
Secondly, a minority expressed concerns about the timeliness of ChatGPT's knowledge sources, suggesting a more positive experience would result from access to information beyond 2021. Thirdly, the company policies for some of our participants (e.g., Participant 2) forbade using generative AI to generate code or requirements, hence hindering the participant's experience.

\surveyquote{Participant 2}{I can't really ask ChatGPT to help me analyze requirements since I am not allowed to share that information outside my company.}

In some cases, we also notice that participants seemed to be unsure of what exactly they are allowed to use ChatGPT for, often opting for a cautious approach. However, some participants seem surprisingly unconcerned with intellectual property rights issues, freely sharing production code with ChatGPT. We were not able to determine whether this was covered by these participants' company policy, or if the participants were unaware of or unwilling to follow the policy. Our study was conducted before OpenAI's announcement of ChatGPT Enterprise (August, 2023) which is a service to provide privacy, security, and deployment for organisations sending their data to ChatGPT.\footnote{\url{https://openai.com/blog/introducing-chatgpt-enterprise}} Consequently, we could not assess how the data privacy measures provided by this service might influence users' experiences when interacting with ChatGPT.

\begin{keyfinding}
\textbf{Findings:} Legal questions and concerns regarding company policies influence the experience of ChatGPT users, both in terms of uncertainty whether they are allowed to use the tool for a specific use case as well as in terms of not being able to provide sufficient context or using it for specific tasks. Additionally, some participants are concerned about a stale knowledge base of the tool.
\end{keyfinding}

\subsection{Personal Experience}
\label{sec:outcome}

Here, we analyse our exit survey results, namely, (i) how useful participants perceive ChatGPT, and (ii) what level of trust they place in it. 
The \textbf{usefulness} of ChatGPT reflects whether participants' needs and expectations were met during their interactions with the tool.
As participants were using ChatGPT during their normal work, it was infeasible to ask them for an assessment of individual dialogues. Instead, we ask for a general evaluation as part of the exit survey at the end of the study period. Therein, our participants used a Likert scale from 1 (not helpful at all) to 6 (extremely helpful) to assess how helpful they found ChatGPT in various aspects, along with open-ended questions to allow them to provide more detailed descriptions.

\begin{figure*}[t]
    \centering
    \includegraphics[width=0.85\linewidth]{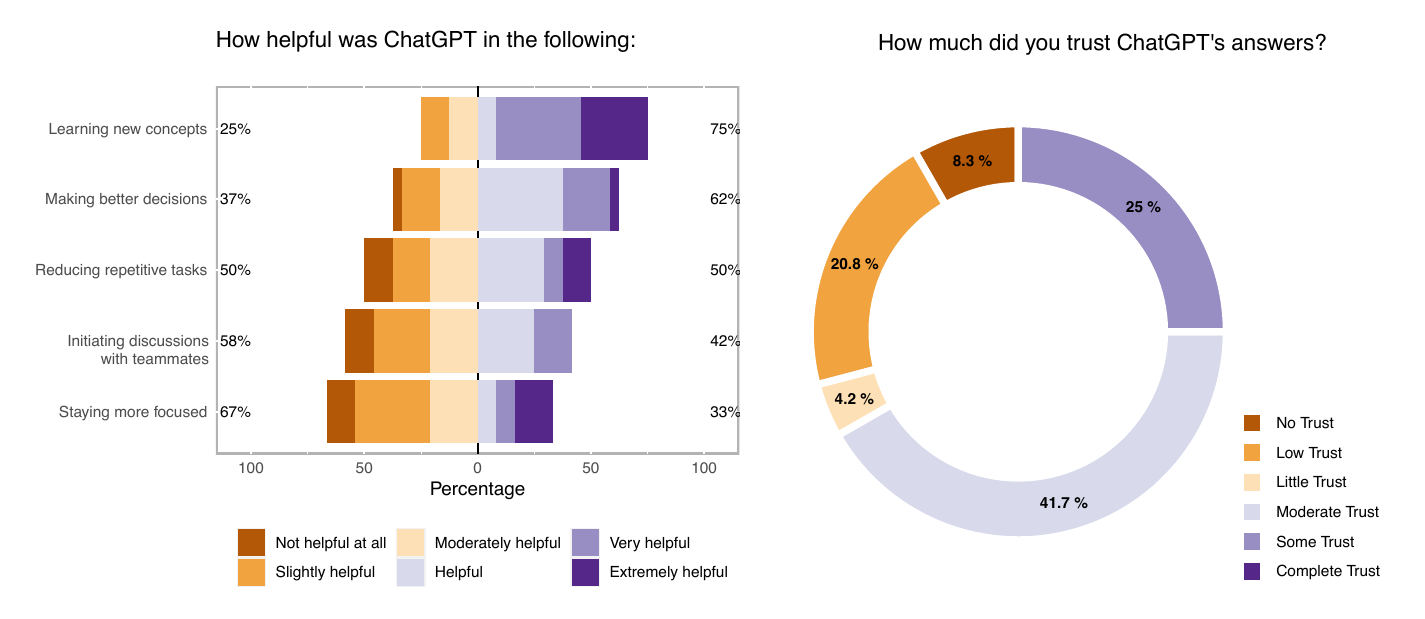}
    \caption{Plots showing how the 23 participants reported ChatGPT's usefulness (left) and trust in its answer (right).}
    \label{fig:usefull_results}
    \vspace{-10pt}
\end{figure*}

Figure~\ref{fig:usefull_results} (left-hand side) shows a relatively high level of usefulness for ChatGPT in terms of learning new concepts and making better decisions. In contrast, participants did not find ChatGPT helpful for initiating discussions with their teams or maintaining focus during work. Meanwhile, there is an even split between participants who believe that ChatGPT is useful for reducing repetitive tasks. A closer look at their dialogues' intent and purpose revealed some differences.

We found that participants who use ChatGPT primarily with an expert consultation purpose to explain and interpret an artifact (e.g., an error message, or source code) found ChatGPT helpful in reducing repetitive tasks, but not in staying focused. These participants found the tool more helpful than users who used it for other information-seeking purposes. There was no noticeable trend among participants who use it for training regarding the reduction of repetitive tasks, but we observed that all these participants found ChatGPT helpful in learning new concepts and staying focused.
\vspace{4px}
\surveyquote{Participant 18}{It is a good tutor and can help with knowledge}

\surveyquote{Participant 1}{Helped me a lot, especially when I needed to learn how to use [a technology]}

On the other hand, participants who use ChatGPT with an artifact manipulation intent found ChatGPT helpful in reducing repetitive tasks that involved generating simple artifacts or blueprints (i.e., a first structure, Participant 5), rather than generating complex artifacts that involve many constraints and components (Participant 13). \newtext{These results align with the findings by Waseem et al. }\cite{waseem2023using}\newtext{, who suggest that ChatGPT becomes more challenging to use with software-related tasks that require complex decision-making.}

\surveyquote{Participant 5}{For code purposes, it was great to get a first structure}

\surveyquote{Participant 13}{Topics that are hard to discuss with ChatGPT are the really complex programming issues, for example, if you have a giant backend system where you need a lot of context to make an accurate determination of what to do, that's very challenging...}

Regarding \textbf{trust} (results shown in Figure \ref{fig:usefull_results}, right-hand side), we note that there is a surprisingly high level of trust in ChatGPT responses --- approximately two thirds of participants place high or moderate trust in the generated artifacts and information that ChatGPT provides, but none of our participants trusts it entirely. A detailed analysis of the dialogues and answers from participants with low trust reveals that lack of sources is the main reason detrimental to trust in ChatGPT.
Similar findings were reported by Ross et al.~\cite{ross2023programmer}, where people who used Codex models wished for a source to be provided along with the solutions.
ChatGPT by default does not provide a source and, if asked to provide one, frequently hallucinates (makes up a non-existent source). This is made even worse by the fact that ChatGPT often shows high confidence even when hallucinating. This led to some participants deciding not to use it further for work-related activities. Others suggest that generative AI should indicate the confidence level of an answer for more transparency.
Interestingly, none of the participants who use the tool for training seem to be particularly concerned about the lack of sources. Generally, this population places a lot of trust in the correctness of ChatGPT's output.

We also noticed that practitioners who employed ChatGPT for generating artifacts tended to have lower levels of trust compared to those using it for brainstorming or modifying artifacts. A potential reason for this may be that queries with a artifact generation purpose often lacked sufficient context, and mistakes are easily visible (e.g., if the artifact does not compile or work as intended).  Additionally, when tasks were more technical, complex, and company-specific, our participants were less inclined to place trust in the results, as exemplified by Participant 2. Participants were cognizant that ChatGPT might struggle to provide accurate responses in domains requiring specialized knowledge, particularly when a software company utilized technologies with limited public documentation. In such cases, the likelihood of ChatGPT offering erroneous solutions increased, thereby impacting trust and confidence in its answers \cite{dzindolet2003trust, vries2003errors}. 

\surveyquote{Participant 2}{I would also not put too much trust in the answers to such complex questions}

Despite the lack of trust, some participants still found value in using ChatGPT, even if it meant spending extra time on fact-checking. Particularly, they saw value in a tool providing a starting point, even if they are aware that manual checking and correcting will be required. 

\begin{keyfinding}
    \textbf{Findings:} Participants found ChatGPT helpful for learning new concepts and making decisions. It can also help reduce repetitive tasks, but only if those tasks are sufficiently simple or context-independent. About two thirds of our study population puts high trust in ChatGPT. The 33\% who trusted it less were often concerned about the lack of sources. Some participants found it helpful to use ChatGPT's answer as a starting point for a solution, even if it meant having to fact check and/or improve it.
\end{keyfinding}

\section{Discussion}
\label{sec:discussion}

We now discuss the main implications and lessons learned from our study, followed by a discussion of threats to validity.

\subsection{Implications}

\noindent \textbf{I1: ChatGPT is more frequently used for receiving guidance and training than for directly generating code.}\\
Despite significant public attention on the fact that LLMs can generate working code, we observed that only around a third of dialogues were actually related to artifact manipulation. In the majority of conversations, the participant was interested in guidance or training \emph{how} to solve a task, not get it automated directly. Note that this finding may be specific to ChatGPT --- tools that are integrated into development environments may well be used more frequently for artifact manipulation (this also allows these tools to provide better context as a ``hidden'' part of the prompt, consequently improving the quality of the generated code). Still, we argue that future studies on LLM usage in software engineering need to take into account that developers use them for more than just generating code.\\

\noindent\textbf{I2: ChatGPT can be helpful in all phases of the software development lifecycle, not just for implementation.}\\
Much of the public debate around ChatGPT for engineers centers around their potential for helping with programming, more specifically, generating code. In our study, we have also seen that most usage was indeed in the implementation phase. However, some of our participants also used the tool extensively and successfully in other phases of the lifecycle, e.g., to improve their process or generate requirements or test cases. In general, we conclude that it may be useful to think more creatively about the types of tasks that an LLM can be used for successfully. Some participants who did not find ChatGPT useful were struggling with identifying use cases for their work, and one reason for this may well be that they had too limited perception in their mind about what kinds of tasks they could use it for.\\

\noindent\textbf{I3: Engineers can, and often do, use ChatGPT despite being aware that results need to be double-checked.}\\
Much like seeking help from a colleague at work, our participants often turn to ChatGPT for assistance, recognizing that, in both cases, absolute accuracy cannot always be guaranteed. Even participants who had fairly low trust in ChatGPT's answers in some cases kept using it. However, these participants also made sure to carefully check any suggestions. 
Generally speaking, we saw that \emph{awareness} was an important factor --- participants did not mind using ChatGPT as long as they could tell how ChatGPT generated its answers, allowing them to judge how plausible they were. We argue that future LLMs should provide more transparency with regard to confidence and remove some of the current guesswork of deciding whether any specific LLM output is potentially a hallucination. One way to do that is displaying the confidence score of ChatGPT's response, which has been shown previously to increase the trust in automated tools that affect decision making \cite{antifakos2005towards, zhang2020confidence}. \\

\noindent\textbf{I4: ChatGPT can be useful even for creative tasks, not just to solve clearly-specified problems.}\\
While many engineers used ChatGPT to directly perform straightforward work tasks, we have also seen many cases where participants were rather looking for inspiration or solution alternatives. Participants who used ChatGPT for brainstorming expressed satisfaction with the output. This is in line with the experiences made by others, indicating that listing plausible partial solutions to a complex problem may, in fact, be an ideal use case for LLMs.\footnote{\url{https://cacm.acm.org/blogcacm/can-llms-really-reason-and-plan/}}
We note that using an LLM for creative tasks is possible even understanding that LLM solutions will never be truly ``novel'', as any answer will still be based on training data. However, as suggested by Koivisto et al. \cite{koivisto2023grassini}, even this recombination of existing knowledge can still lead to useful, creative results.\\

\noindent\textbf{I5: ChatGPT can, in some ways, hurt developer productivity.}\\
Based on existing productivity frameworks for developers \cite{storey2016disrupting, forsgren2021space}, we found that although ChatGPT can be helpful for learning or automating routine tasks, it can actually reduce productivity in some dimensions: \textit{(i)} its usage reduces team communication, as questions better asked to a colleague are sometimes directed to the chatbot, and \textit{(ii)} reducing participant focus. We have observed these problems in different facets of our study. For example, in some cases, an overconfident AI response may have given a participant the feeling that further discussion with the rest of the team is not required. Additionally, we observed that in some cases participants spent an extraordinary amount with tweaking prompts, trying to get ChatGPT to generate perfectly working code (rather than simply taking a slightly defect one and fixing it). To some extent, this may be an effect of our limited study window, with many participants clearly still in an ``experimenting with AI'' phase. However, this result also aligns with previous work by Vaithilingam et al. \cite{vaithilingam2022usability}, who found that using Co-Pilot did not improve the effectiveness or efficiency of developers in many cases.

\subsection{Threats to validity}
The \textit{external validity} of our research has its limitations. Firstly, our sample of participants may not be fully representative of software engineers population, due to its limited sample size and geographical region \newtext{(Europe).} However, \newtext{we ensured that the companies we selected are from three non-neighboring countries. We also}have diversity in size and domain of software organisations.

\newtext{Secondly, the nature of our study involving participants sharing their logs can introduce bias from participants removing complete or parts of their chat logs. To mitigate this, we instructed the participants to send us their chat logs in the form of web archive files. This format does not allow any alteration or removal of parts of the chats. We also ensured full anonymity to encourage the participants to share all their logs.
}

Thirdly, we conducted our study with a single LLM-based chatbot (ChatGPT). We acknowledge that the theoretical framework described in Section~\ref{sec:results} will allow researchers to capture the experience of users of some, but not all, other AI developer tools. For example, for a tool such as Co-Pilot, which provides a more integrated experience with less focus on explicit prompts, the internal and external factors driving the experience may be fairly different. Therefore, we encourage researchers to experiment with and expand the proposed framework to include other facets and dimensions of chatbot usage. 
\newtext{Similarly, the version of ChatGPT that we used was powered by GPT-3.5 since it was accessible to all participants. Given that newer versions (e.g., GPT-4) may show increased performance in some areas,
we hypothesize that such models can strengthen our findings about the use cases where ChatGPT was helpful and vice versa for not useful use cases. For example, GPT-4 can add more value to use cases that we found to be useful using GPT-3.5 (such as brainstorming), and hence increase the usefulness. While in some cases where GPT-3.5 was not helpful, such as, generating complex artifacts, using a better-performing model can overcome these challenges and provide more helpful responses. 
To further investigate this, we suggest using our framework to explore the impact of the model version on different use cases and other factors that we illustrate in Figure }\ref{fig:framework}.

Regarding \textit{internal validity}, the classification of our dialogues includes a degree of subjectivity inherent to an inductive qualitative study. We mitigate this limitation by creating the classification schema described in Section~\ref{sec:method}. The proposed schema expanded on existing classifications and was refined over three rounds of discussion and classification involving many of the authors.

In terms of \textit{construct validity}, our proposed dialogue categories cover only a set of purposes that can be expanded as we gather more usage-focused evidence on chatbots. We argue that the three proposed levels are sufficient for our scope as they can be mapped to distinct types of dialogues with corresponding high-level intents.
Regarding the construction of our survey, the Likert scales can limit the depth or accuracy in the participants' responses. We complement those scales with open-ended questions to obtain a better picture of the different aspects of the participants' experience and interactions. Furthermore, the choice of questions was a byproduct of the feedback received during our pilot study, i.e., before the survey and material were sent to participants.

\section{Conclusion}
We reported on the results of an observational, primarily qualitative, study on the usage of ChatGPT in a software engineering context. We observed participants for one week, and analysed their dialogues with ChatGPT as well as their overall impressions as expressed in an exit survey. We propose a theoretical framework for how high-level purpose of the dialogue, internal factors such as participant personality, and external factors such as company policy, together shape the experience (in terms of perceived usefulness and trust) of software engineers using an LLM such as ChatGPT. 

We envision that our framework can be used by future research to further the academic discussion on LLM usage by practitioners. Future studies should take into account that a majority of ChatGPT usage is not to generate directly-usable code, and that LLMs are also used for software engineering tasks outside of coding. We believe that particularly using LLMs for learning and training is currently under-researched, and future empirical work on this aspect will be required. Finally, more research on detailed prompt engineering for software engineering tasks is required. In our study, we saw engineers habitually repeat patterns they observed on the Internet, without much evidence that these are indeed the most effective ways to utilize an LLM as a software engineer.

\section*{Data Availability}
The files used to generate plots and tables are available in our anonymised re-analysis package in Zenodo \cite{khojah2023replication}. The package includes the exit survey questionnaire, along with the CSV files with the classification of dialogues and Likert answers from our participants. We cannot share the files with the chats or the survey's open-ended answers because they might break the anonymity of our participants or include company-specific information which is protected under Non-Disclosure Agreements (NDA) with our industry partners.

\section*{Acknowledgment}
This work was partially supported by the Wallenberg AI, Autonomous Systems and Software Program (WASP) funded by the Knut and Alice Wallenberg Foundation. It was also partially supported by AGRARSENSE, a project funded by the Chips JU and its members, including the top-up funding by Sweden, Czechia, Finland, Ireland, Italy, Latvia, Netherlands, Norway, Poland and Spain (Grant Agreement No.101095835).

\bibliographystyle{ieeetr}
\bibliography{bibliography.bib}

\begin{thebibliography}{10}

\bibitem{santhanam2022bots}
S.~Santhanam, T.~Hecking, A.~Schreiber, and S.~Wagner, ``Bots in software
  engineering: a systematic mapping study,'' {\em PeerJ Computer Science},
  vol.~8, p.~866, 2022.

\bibitem{erlenhov2019current}
L.~Erlenhov, F.~G. de~Oliveira~Neto, R.~Scandariato, and P.~Leitner, ``Current
  and future bots in software development,'' in {\em 2019 IEEE/ACM 1st
  International Workshop on Bots in Software Engineering (BotSE)}, pp.~7--11,
  IEEE, 2019.

\bibitem{brown2020language}
T.~B. Brown, B.~Mann, N.~Ryder, M.~Subbiah, J.~Kaplan, P.~Dhariwal,
  A.~Neelakantan, P.~Shyam, G.~Sastry, A.~Askell, S.~Agarwal, A.~Herbert-Voss,
  G.~Krueger, T.~Henighan, R.~Child, A.~Ramesh, D.~M. Ziegler, J.~Wu,
  C.~Winter, C.~Hesse, M.~Chen, E.~Sigler, M.~Litwin, S.~Gray, B.~Chess,
  J.~Clark, C.~Berner, S.~McCandlish, A.~Radford, I.~Sutskever, and D.~Amodei,
  ``Language models are few-shot learners,'' in {\em Proceedings of the 34th
  International Conference on Neural Information Processing Systems}, NIPS '20,
  (Red Hook, NY, USA), Curran Associates Inc., 2020.

\bibitem{chen2021evaluating}
M.~Chen, J.~Tworek, H.~Jun, Q.~Yuan, H.~P. d.~O. Pinto, J.~Kaplan, H.~Edwards,
  Y.~Burda, N.~Joseph, G.~Brockman, {\em et~al.}, ``Evaluating large language
  models trained on code,'' 2021.

\bibitem{fraiwan2023review}
M.~Fraiwan and N.~Khasawneh, ``A review of chatgpt applications in education,
  marketing, software engineering, and healthcare: Benefits, drawbacks, and
  research directions,'' 2023.

\bibitem{mastropaolo2023robustness}
A.~Mastropaolo, L.~Pascarella, E.~Guglielmi, M.~Ciniselli, S.~Scalabrino,
  R.~Oliveto, and G.~Bavota, ``On the robustness of code generation techniques:
  An empirical study on github copilot,'' in {\em Proceedings of the 45th
  International Conference on Software Engineering}, ICSE '23, (Piscataway,
  NJ), p.~2149–2160, IEEE Press, 2023.

\bibitem{nguyen2022empirical}
N.~Nguyen and S.~Nadi, ``An empirical evaluation of github copilot's code
  suggestions,'' in {\em Proceedings of the 19th International Conference on
  Mining Software Repositories}, (New York, NY, USA), pp.~1--5, Association for
  Computing Machinery, 2022.

\bibitem{qian2023communicative}
C.~Qian, X.~Cong, C.~Yang, W.~Chen, Y.~Su, J.~Xu, Z.~Liu, and M.~Sun,
  ``Communicative agents for software development,'' 07 2023.

\bibitem{ezzini2023ai}
S.~Ezzini, S.~Abualhaija, C.~Arora, and M.~Sabetzadeh, ``Ai-based question
  answering assistance for analyzing natural-language requirements,'' in {\em
  Proceedings of the 45th International Conference on Software Engineering},
  ICSE '23, (Piscataway, NJ), p.~1277–1289, IEEE Press, 2023.

\bibitem{lemieux2023codamosa}
C.~Lemieux, J.~P. Inala, S.~K. Lahiri, and S.~Sen, ``Codamosa: Escaping
  coverage plateaus in test generation with pre-trained large language
  models,'' in {\em Proceedings of the 45th International Conference on
  Software Engineering}, ICSE '23, (Piscataway, NJ), p.~919–931, IEEE Press,
  2023.

\bibitem{tufano2023automating}
R.~Tufano, L.~Pascarella, and G.~Bavota, ``Automating code-related tasks
  through transformers: The impact of pre-training,'' 2023.

\bibitem{wood2018detecting}
A.~Wood, P.~Rodeghero, A.~Armaly, and C.~McMillan, ``Detecting speech act types
  in developer question/answer conversations during bug repair,'' in {\em
  Proceedings of the 2018 26th ACM joint meeting on european software
  engineering conference and symposium on the foundations of software
  engineering}, (New York, NY, USA), pp.~491--502, Association for Computing
  Machinery, 2018.

\bibitem{surameery2023use}
N.~M.~S. Surameery and M.~Y. Shakor, ``Use chat gpt to solve programming
  bugs,'' {\em International Journal of Information Technology \& Computer
  Engineering (IJITC) ISSN: 2455-5290}, vol.~3, no.~01, pp.~17--22, 2023.

\bibitem{forsgren2021space}
N.~Forsgren, M.-A. Storey, C.~Maddila, T.~Zimmermann, B.~Houck, and J.~Butler,
  ``The space of developer productivity: There's more to it than you think.,''
  {\em Queue}, vol.~19, no.~1, pp.~20--48, 2021.

\bibitem{storey2016disrupting}
M.-A. Storey and A.~Zagalsky, ``Disrupting developer productivity one bot at a
  time,'' in {\em Proceedings of the 2016 24th ACM SIGSOFT International
  Symposium on Foundations of Software Engineering}, FSE 2016, (New York, NY,
  USA), p.~928–931, Association for Computing Machinery, 2016.

\bibitem{kimani2019conversational}
E.~Kimani, K.~Rowan, D.~McDuff, M.~Czerwinski, and G.~Mark, ``A conversational
  agent in support of productivity and wellbeing at work,'' in {\em 2019 8th
  international conference on affective computing and intelligent interaction
  (ACII)}, pp.~1--7, IEEE, 2019.

\bibitem{ahmad2023towards}
A.~Ahmad, M.~Waseem, P.~Liang, M.~Fahmideh, M.~S. Aktar, and T.~Mikkonen,
  ``Towards human-bot collaborative software architecting with chatgpt,'' in
  {\em Proceedings of the 27th International Conference on Evaluation and
  Assessment in Software Engineering}, (New York, NY, USA), pp.~279--285,
  Association for Computing Machinery, 2023.

\bibitem{peng2023impact}
S.~Peng, E.~Kalliamvakou, P.~Cihon, and M.~Demirer, ``The impact of ai on
  developer productivity: Evidence from github copilot,'' 2023.

\bibitem{waseem2023using}
M.~Waseem, T.~Das, A.~Ahmad, M.~Fehmideh, P.~Liang, and T.~Mikkonen, ``Using
  chatgpt throughout the software development life cycle by novice
  developers,'' 2023.

\bibitem{weisz2022better}
J.~D. Weisz, M.~Muller, S.~I. Ross, F.~Martinez, S.~Houde, M.~Agarwal,
  K.~Talamadupula, and J.~T. Richards, ``Better together? an evaluation of
  ai-supported code translation,'' in {\em 27th International Conference on
  Intelligent User Interfaces}, (New York, NY, USA), pp.~369--391, Association
  for Computing Machinery, 2022.

\bibitem{barke2023grounded}
S.~Barke, M.~B. James, and N.~Polikarpova, ``Grounded copilot: How programmers
  interact with code-generating models,'' {\em Proceedings of the ACM on
  Programming Languages}, vol.~7, no.~OOPSLA1, pp.~85--111, 2023.

\bibitem{euro-org-size}
E.~Commision, ``Internal market, industry, entrepreneurship and smes,'' 2021.
\newblock Accessed on May 10, 2024.

\bibitem{walton2009burden}
D.~Walton, ``Burden of proof in deliberation dialogs,'' in {\em Argumentation
  in Multi-Agent Systems}, (Berlin, Heidelberg), pp.~1--22, Springer, Springer
  Berlin Heidelberg, 2010.

\bibitem{daSilva2015being}
C.~da~Silva~Cintra and R.~A. Bittencourt, ``Being a pbl teacher in computer
  engineering: An interpretative phenomenological analysis,'' in {\em 2015 IEEE
  Frontiers in Education Conference (FIE)}, pp.~1--8, IEEE, 2015.

\bibitem{eatough2017interpretative}
V.~Eatough and J.~A. Smith, ``Interpretative phenomenological analysis,'' {\em
  The Sage handbook of qualitative research in psychology}, pp.~193--209, 2017.

\bibitem{melo2020exploring}
G.~M. dos Santos, E.~Law, P.~S.~C. Alencar, and D.~Cowan, ``Exploring
  context-aware conversational agents in software development,'' {\em CoRR},
  vol.~abs/2006.02370, 2020.

\bibitem{erlenhov2020empirical}
L.~Erlenhov, F.~G. de~Oliveira~Neto, and P.~Leitner, ``An empirical study of
  bots in software development: characteristics and challenges from a
  practitioner’s perspective,'' in {\em Proceedings of the 28th ACM Joint
  Meeting on European Software Engineering Conference and Symposium on the
  Foundations of Software Engineering}, ESEC/FSE 2020, (New York, NY, USA),
  p.~445–455, Association for Computing Machinery, 2020.

\bibitem{ross2023programmer}
S.~I. Ross, F.~Martinez, S.~Houde, M.~Muller, and J.~D. Weisz, ``The
  programmer’s assistant: Conversational interaction with a large language
  model for software development,'' in {\em Proceedings of the 28th
  International Conference on Intelligent User Interfaces}, (New York, NY,
  USA), pp.~491--514, Association for Computing Machinery, 2023.

\bibitem{dzindolet2003trust}
M.~T. Dzindolet, S.~A. Peterson, R.~A. Pomranky, L.~G. Pierce, and H.~P. Beck,
  ``The role of trust in automation reliance,'' {\em Int. J. Hum.-Comput.
  Stud.}, vol.~58, p.~697–718, jun 2003.

\bibitem{vries2003errors}
P.~de~Vries, C.~Midden, and D.~Bouwhuis, ``The effects of errors on system
  trust, self-confidence, and the allocation of control in route planning,''
  {\em Int. J. Hum.-Comput. Stud.}, vol.~58, p.~719–735, jun 2003.

\bibitem{antifakos2005towards}
S.~Antifakos, N.~Kern, B.~Schiele, and A.~Schwaninger, ``Towards improving
  trust in context-aware systems by displaying system confidence,'' in {\em
  Proceedings of the 7th international conference on Human computer interaction
  with mobile devices \& services}, (New York, NY, USA), pp.~9--14, Association
  for Computing Machinery, 2005.

\bibitem{zhang2020confidence}
Y.~Zhang, Q.~V. Liao, and R.~K.~E. Bellamy, ``Effect of confidence and
  explanation on accuracy and trust calibration in ai-assisted decision
  making,'' in {\em Proceedings of the 2020 Conference on Fairness,
  Accountability, and Transparency}, FAT* '20, (New York, NY, USA),
  p.~295–305, Association for Computing Machinery, 2020.

\bibitem{koivisto2023grassini}
M.~Koivisto and S.~Grassini, ``Best humans still outperform artificial
  intelligence in a creative divergent thinking task,'' {\em Scientific
  Reports}, vol.~13, p.~13601, Sep 2023.

\bibitem{vaithilingam2022usability}
P.~Vaithilingam, T.~Zhang, and E.~L. Glassman, ``Expectation vs. experience:
  Evaluating the usability of code generation tools powered by large language
  models,'' in {\em Extended Abstracts of the 2022 CHI Conference on Human
  Factors in Computing Systems}, CHI EA '22, (New York, NY, USA), Association
  for Computing Machinery, 2022.

\bibitem{khojah2023replication}
R.~Khojah, M.~Mohamad, P.~Leitner, and F.~G. de~Oliveira~Neto, ``{Package for
  An Observational Study of ChatGPT Usage in Software Engineering Practice},''
  Sept. 2023.

\end{thebibliography}

\end{document}